\documentclass[twocolumn,aps,superscriptaddress]{revtex4}
\usepackage{epsfig}

\newcommand{\be}{\begin{equation}}
\newcommand{\ee}{\end{equation}}
\newcommand{\bea}{\begin{eqnarray}}
\newcommand{\eea}{\end{eqnarray}}

\newcommand{\bc}{\begin{center}}
\newcommand{\ec}{\end{center}}

\newcommand{\forget}[1]{}

\begin{document}

\preprint{}
\title{Detecting entanglement with partial state information}
\author{Federico M. Spedalieri}
\email{fspedali@isi.edu}
\affiliation{Information Sciences Institute, and}
\affiliation{Center for Quantum Information Sciences and Technology, Viterbi
School of Engineering, University of Southern California}

\date{\today}

\begin{abstract}

We introduce a sequence of numerical tests that can determine the entanglement or separability of a
state even when there is not enough information to completely determine its density matrix. Given 
partial information about the state in the form of linear constraints on the density matrix, 
the sequence of tests can prove that either all states satisfying the constraints are entangled, or  there
is at least one separable state that satisfies them. The algorithm works even if the values of the
constraints are only known to fall in a certain range. If the states are entangled, an entanglement witness
is constructed and lower bounds on entanglement measures and related quantities are provided;
 if a separable state satisfies the constraints, a separable decomposition is provided to certify
this fact.

\end{abstract}

\pacs{}

\maketitle

\section{Introduction}

Entanglement is one of the central features in quantum information processing (QIP). It has been identified
as a key ingredient in many useful QIP tasks such as teleportation, quantum key distribution, superdense
coding and quantum computation~\cite{nielsen2000}. It is also remarkable that even when a complete
description of a quantum state is given (in the form of a density matrix), it can be extremely difficult to computationally decide
whether such state is entangled or not. This is due to the fact that this problem (usually called 
``the separability problem") is known to be NP-Hard~\cite{gurvits2003a}. The problem is even more 
difficult when we consider experimental tests of entanglement, since measurements may not provide
a full description of the state, and  when they do (such as in quantum state
tomography~\cite{vogel1989a}) the reconstructed density matrix may be unphysical (i.e., not positive
semidefinite (PSD)).

A key problem with important practical applications is to determine the entanglement characteristics
of a state when only a limited amount of information is available. If this information comes from 
measuring a set of observables, it takes the form of a set of linear constraints on the elements of
the density matrix. In this article we will introduce a sequence of numerical tests that can decide whether all states 
that satisfy a set of linear constraints are entangled, or if there is at least one separable state
that satisfies those same constraints. The approach is based on an extension of the PPT Symmetric Extension
(PPTSE) criterion~\cite{doherty2002a} and its dual introduced in \cite{navascues2009a}. If the states are shown to be  entangled,
the algorithm constructs an entanglement witness (EW) that certifies this fact for all such states and
such a witness can be used to provide lower bounds on certain entanglement measures and related quantities.
If a separable state satisfying the constraints exists, the algorithm finds it and provides
a separable decomposition as a proof.

The paper is organized as follows: in Section \ref{PPTSE} we review the PPTSE criterion and
its dual; Section \ref{partial} shows how to extend these two criteria to the case where only partial
information about the state  $\rho$ is available in the form of a set of linear constraints;
Section \ref{witness} shows how to construct and entanglement witness if all states
satisfying the constraints are shown to be entangled; if the state is
shown to be entangled, Section \ref{entmeasures} 
provides lower bounds on entanglement measures and other related quantities; Section \ref{example}
shows an example of the application of this technique; Section \ref{features} discusses
some basic features of the approach and our conclusions are presented in Section \ref{conclusions}.

\section{The PPT Symmetric Extension criterion and its dual}
\label{PPTSE}

To determine the entanglement or separability of a state $\rho$ we will use the PPTSE criterion~\cite{doherty2003d} and a dual approach
introduced by Navascu\'es et al.,\cite{navascues2009a}.  When used together, these two criteria can conclusively determine
if a state is separable or entangled in a finite number of steps (however, the number of steps and the computational
resources required to implement them can be arbitrarily high for some states). Let us start with some definitions. If  $\rho$ is a 
state in ${\cal H}_A  \otimes {\cal H}_B$, we will call $\tilde\rho$ in  ${\cal H}_A^{\otimes k} \otimes {\cal H}_B$
a PPT symmetric extension of $\rho$ to $k$ copies of subsystem $A$ if:
$(i) \  \rho = {\mathrm{Tr}}_{A^{k-1}} [\tilde\rho]$, $(ii) \ \tilde\rho$ is symmetric under exchanges of copies of subsystem $A$, and 
$(iii) \ \tilde\rho$ has positive partial transposes for any bipartite arrangement of the subsystems $A$ and $B$. The key point is
that, since any separable state
in ${\mathcal{H}}_{A} \otimes {\mathcal{H}}_{B}$ can be written as $\rho =\sum p_{i}|\psi _{i}\rangle \langle \psi _{i}|\otimes |\phi
_{i}\rangle \langle \phi _{i}|$, it trivially has such an extension given by $\tilde\rho =\sum p_{i}|\psi _{i}\rangle \langle \psi _{i}| ^{\otimes k} 
\otimes |\phi_{i}\rangle \langle \phi _{i}|$.
For each value of $k$, the non-existence of a PPTSE provides a sufficient (but not necessary) condition for entanglement. In the limit
$k \to \infty$ the condition becomes necessary. The practical value of this approach is that searching for such extensions
or proving their impossibility can be cast as a semidefinite program (SDP).

An SDP is a type of convex optimization problem that has a broad range of applications
and has been widely applied in quantum information. An SDP has both a primal and a dual form. 
A typical SDP in its primal form reads
\begin{eqnarray}
\label{sdp}
{\mathrm{minimize}} &\ \  c^T {\bf x}  \nonumber \\
{\mathrm{ subject \ to}} &\ \  F_0 + \sum_i x_i F_i \succeq 0, 
\end{eqnarray}
where $c$ is a given vector, ${\bf x} = (x_1,\ldots,x_n)$, and
$ F_0$ and $F_i$ are some fixed Hermitian matrices. The inequality in the second line
means that the affine combination of the $F$ matrices must be positive semidefinite. The 
minimization is performed over the vector ${\bf x}$, whose components
are the variables of the problem. The dual of this SDP takes the form
\begin{eqnarray}
\label{dual2}
{\mathrm{maximize}} &\ \ -\text{Tr} [F_0 Z]  \nonumber \\
{\mathrm{subject \ to}} &\ \  Z \succeq 0, \nonumber \\
                       &\ \ \text{Tr} [F_i Z] = c_i,
\end{eqnarray}
where the dual variables are the components of the matrix $Z$.

Each test in the PPTSE hierarchy provides a sufficient but not necessary condition for entanglement. The hierarchy
is complete in the limit: any entangled state is guaranteed to be detected by one of the tests. But a separable
state will pass all tests, leading to a non-terminating algorithm.
Fortunately, a dual approach was developed by Navascu\'es et al.,\cite{navascues2009a}, that applies a sequence
of tests that  can certify separability in a finite number of steps (although that number can be very high for some states).
Geometrically, the PPTSE hierarchy of tests works by monotonically approximating the cone of separable states from the outside
with a sequence of cones associated with states having PPT symmetric extensions to a certain number of copies
of one of the subsystems. The dual
approach in \cite{navascues2009a} constructs a similar approximation to the cone of separable states, but from the inside:
it provides sufficient (but not necessary) conditions for separability. By interleaving the two sequences of tests we can,
in a finite number of steps,
determine if a state is entangled (and give an entanglement witness as a proof), or separable (and provide
an explicit separable decomposition). We will now briefly describe the test in \cite{navascues2009a}.

Let ${\cal S}^N_p$ be the set of states in  ${\cal H}_A \otimes {\cal H}_B$ that have a PPT symmetric extension to $N$ copies
of $A$. In \cite{navascues2009b} it was shown that a small perturbation in $ {\cal H}_B$ makes these states separable. More precisely
we have that
$ {\tilde{\cal S}}_p^N \equiv \{ (1-\epsilon_N) \omega_{AB} + \epsilon_N \, \omega_A \otimes \frac{\mathbf{1}_B}{d_B} : 
\omega_{AB} \in {\cal S}^N_p \}$
satisfies ${\tilde{\cal S}}_p^N \subset {\cal S}$, for all $N$, where ${\cal S}$ is the set of separable states in ${\cal H}_A \otimes {\cal H}_B$ ,
$\omega_A = \mathrm{Tr}_B [\omega_{AB}]$,
and 
$ \epsilon_N \equiv d_B /(2(d_B -1))  \mathrm{ min} \{1-x : P_{\lfloor N/2 \rfloor +1}^{(d_B-2,N \, \mathrm{mod} \, 2)} (x) =0 \}$,
with $P_n^{(\alpha,\beta)}(x)$ the Jacobi polynomials. Since ${\tilde{\cal S}}_p^N \to {\cal S}_p^N$ for $N$ going to infinity, and
${\cal S} \subset{\cal S}_p^N $ for all $N$, we have that ${\tilde{\cal S}}_p^N \to {\cal S} \ (N \to \infty)$. 
This result can be easily transformed into a SDP like (\ref{sdp}) that tests
if a given state is separable~\cite{navascues2009a}. If that is the case, the output of the SDP can be used
to construct an explicit separable decomposition (the details of this construction can be found in \cite{navascues2009b}).

\section{Entanglement testing with partial state information}
\label{partial}

The PPTSE criterion and its dual discussed above require as input the complete density matrix,
and so cannot be applied directly if we only have access to partial information about the state.
We will now show that they can be extended so that they can be applied in this more general case.

Consider a situation in which we are given partial information of the state of a quantum system
in the form of $L$ linear constraints on the elements of its density matrix 
\be
\label{linear constraints}
\mathrm{Tr}[\rho M_l ] = m_l, \ \ \  l=1,\ldots,L
\ee
where the operators $M_l$ are arbitrary. Our goal is to determine if all the states satisfying these
constraints are entangled, or if there is a separable state that satisfies them. 
The constraints in (\ref{linear constraints}) are nothing but a linear system of equations for the
elements of the density matrix $\rho$. If this system is incompatible it means that these constraints do not
describe a physical state. If the system is invertible, then the Hermitian matrix $\rho$ can be completely
determined from the equations, and once we have an explicit expression  we can check if it corresponds
to a state (i.e., it is PSD and normalized), and then 
apply the PPTSE criterion and its dual. 
But the situation that is the most interesting (and typically more common) corresponds to the case in which the linear
system (\ref{linear constraints})  is underdetermined, and we do not have enough information to uniquely
define the state. This situation corresponds naturally to being able to measure only the expectation values
of a limited number of observables (the operators $M_l$ in (\ref{linear constraints})
are then Hermitian matrices). We will show that in this case, the linear system defines an affine subspace in the space of Hermitian
matrices, and the PPTSE criterion and its dual can be applied to either prove that all states in that affine subspace
are entangled, or to show that a separable state exists that satisfies (\ref{linear constraints}).

Let us start with the linear system (\ref{linear constraints}). The most general solution $\rho$ of this system can be written
as
\be
\label{gensol}
\rho = \rho^{part} + \sum_{a=1}^{D_K} y_a  \mu^{(a)},
\ee
where $\rho^{part}$ is a particular solution of (\ref{linear constraints}) (i.e., $\mathrm{Tr}[\rho^{part} M_l ] = m_l, \ \ \  l=1,\ldots,L$),
the matrices $\{ \mu^{(a)} \}$ form a basis of the subspace of solutions of the homogeneous system (i.e., $\mathrm{Tr}[ \mu^{(a)} M_l ] = 0, \ \ \  l=1,\ldots,L$),
$D_K$ is the dimension of this subspace, and $y_a$ are real variables. Note that $\rho^{part}$ is just a Hermitian matrix and not necessarily a
state since it need not be PSD or normalized. The question then reduces to whether there are
values of the real variables $y_a$ such that the resulting Hermitian matrix is a normalized, separable state. If there are not,
then all the states of the form (\ref{gensol}), that is all normalized PSD Hermitian matrices satisfying (\ref{linear constraints}) must be entangled.

To test the entanglement of a state of the form (\ref{gensol}), we can apply the PPTSE criterion for any value of $k$.
We will present in detail how this works for $k=2$ (the general case is 
straightforward). So we need to check if, for some values of the variables $y_a$, the resulting matrix
is PSD, normalized and has a PPTSE. Let $\{ \sigma_i^A \}_{i=1}^{d_A^2}, 
\{\sigma_j^B\}_{j=1}^{d_B^2}$ be bases for the spaces of Hermitian
matrices that operate on ${\cal H}_A$ and  ${\cal H}_B$, of
dimensions $d_A$ and $d_B$  respectively, 
such that they satisfy $\mathrm{Tr} [ \sigma_i^X \sigma_j^X] = \alpha_X \delta_{i j}$ and $\mathrm{Tr} [\sigma_i^X]  =  \delta_{i1}$
(where $X$ stands for $A$ or $B$), and $\alpha_X$ is some constant.
Then we can expand $\rho$ in the basis 
$\{ \sigma_i^A \otimes \sigma_j^B \}$, and write
$\rho = \sum_{i j} \rho_{i j} \sigma_i^A \otimes \sigma_j^B$,
with $\rho_{i j}=\alpha_A^{-1}\alpha_B^{-1} {\mathrm{Tr}} [\rho \, \sigma_i^A \otimes 
\sigma_j^B]$. In the same way, we can expand the extension $\tilde\rho$ in
${\cal H}_A^{\otimes 2} \otimes {\cal H}_B $ as
\begin{eqnarray}
\label{rhoext}
\tilde\rho & = & \sum_{\stackrel{ikj}{\tiny{i<k}}}
\tilde\rho_{ikj} \{\sigma_i^A  \otimes \sigma_k^A 
\otimes \sigma_j^B+\sigma_k^A \otimes \sigma_i^A 
\otimes \sigma_j^B \} + \nonumber \\
& & + \sum_{kj}\tilde\rho_{kkj} \,\sigma_k^A  \otimes \sigma_k^A
\otimes \sigma_j^B, 
\end{eqnarray}
where we made explicit use of the swapping symmetry between the
two copies of $A$. To satisfy
the condition that $\tilde\rho$ is an extension of $\rho$, we need to impose $\mathrm{Tr}_A [\tilde\rho] = \rho$.
This implies $\tilde\rho_{i1j} = \rho_{ij}$.
From (\ref{gensol}) we have that 
$\rho_{ij} = \rho^{part}_{ij} + \sum_{a=1}^{D_K} y_a \mu^{(a)}_{ij}$, 
which fixes some of the components of the extension (\ref{rhoext}). We then have
\begin{eqnarray}
\label{rhoextfix}
\tilde\rho & = & \sum_{\stackrel{i j}{\tiny{i>1}}}
\rho^{part}_{ij} \{\sigma_i^A  \otimes \sigma_1^A 
\otimes \sigma_j^B+\sigma_1^A \otimes \sigma_i^A 
\otimes \sigma_j^B \} + \nonumber \\
& & + \sum_{j} \rho^{part}_{1j} \,\sigma_1^A  \otimes \sigma_1^A
\otimes \sigma_j^B + \nonumber \\
& & + \sum_{a=1}^{D_K} y_a \left(  \sum_{\stackrel{i j}{\tiny{i>1}}}
\mu^{(a)}_{ij} \{\sigma_i^A  \otimes \sigma_1^A 
\otimes \sigma_j^B+\sigma_1^A \otimes \sigma_i^A 
\otimes \sigma_j^B \} + \right. \nonumber \\
& & \left. + \sum_{j} \mu^{(a)}_{1j} \,\sigma_1^A  \otimes \sigma_1^A
\otimes \sigma_j^B \} \right)+ \nonumber \\
& & + \sum_{\stackrel{i j k}{\tiny{i>k\geq 2}}}
\tilde\rho_{ikj} \{\sigma_i^A  \otimes \sigma_k^A 
\otimes \sigma_j^B+\sigma_k^A \otimes \sigma_i^A 
\otimes \sigma_j^B \} + \nonumber \\
& & +   \sum_{\stackrel{ j k}{\tiny{k\geq 2}}} \tilde\rho_{kkj} \,\sigma_k^A  \otimes \sigma_k^A
\otimes \sigma_j^B, 
\end{eqnarray}
If we define a vector of variables ${\bf x} = ({\bf{y}},\tilde\rho_{ikj})$ (with $2\leq k\leq i \leq d_A^2, \, 1\leq j \leq d_B^2$),
we can see that the most general form of the extension (\ref{rhoextfix}) has the 
form $ G_0 + \sum_i x_i G_i$, where the expressions for $G_0$ and $G_i$ can
be easily extracted from it.

The first condition we need to impose on this extension is that it represents a state, i.e., that it is PSD and normalized.
The normalization condition can always be assumed to be contained in the set of linear equations (\ref{linear constraints}),
by adding another constraint with $M = \bf{1}$ and expectation value equal to $1$. Requiring that the extension is
PSD means imposing the linear matrix inequality (LMI) $ G_0 + \sum_i x_i G_i \succeq 0$. And finally, imposing
the positivity of the partial transposes requires two more LMIs, namely $ G_0^{T_A} + \sum_i x_i G_i^{T_A}  \succeq 0$
and $ G_0^{T_B} + \sum_i x_i G_i^{T_B}  \succeq 0$ (due to the swapping symmetry, these are the only two independent
partial transposes). We can combine these three LMIs into a single one by defining matrices  
$F_0 = G_0 \oplus G_0^{T_A} \oplus  G_0^{T_B}$ and $F_i = G_i \oplus G_i^{T_A} \oplus  G_i^{T_B}$ ( a block
diagonal matrix is PSD if and only if all of its blocks are PSD). So searching for a PPTSE of a state of the form (\ref{gensol})
corresponds to a SDP of the form (\ref{sdp}) with $c = (0,\ldots,0)$. If there are values of $y_a$ such that (\ref{gensol}) 
is separable, then there must
exist values of $\tilde\rho_{ikj},\ (2\leq k\leq i \leq d_A^2, \, 1\leq j \leq d_B^2)$ such that the SDP is feasible (because
separable states always have PPTSE). But if the SDP is infeasible, it means that there is no set of values $y_a$ for which
the resulting state $\rho$ has a PPTSE, and hence all states of the form (\ref{gensol}) must be entangled.

If the SDP is feasible it means that there is a state with the required PPT symmetric extension that is compatible
with (\ref{linear constraints}). Furthermore, the output of the SDP provides us the values of the variables $\{y_a\}$
of such state, so we can completely determine its density matrix using (\ref{gensol}). However, this does not mean
that this state has to be separable. To determine if this is actually  the case we need to apply the dual criterion of Navascu\'es et al., to
this state. As discussed in Section \ref{PPTSE} this is just another SDP. If the state is proven to be separable, an 
explicit separable decomposition can be constructed~\cite{navascues2009b}.

\subsection{Extension to more general constraints}

We can take this approach a little bit further and consider the case in which the expectation values
of the operators $M_l$ (the right-hand side of (\ref{linear constraints})) are known only approximately.
Assume that instead of (\ref{linear constraints}) we have 
\be
\label{ineqconstr}
m^{min}_l \leq \mathrm{Tr}[\rho M_l ]  \leq m^{max}_l \ \ \  l=1,\ldots,L.
\ee
Now consider the set of matrices $\{ \tau_p : \mathrm{Tr}[\tau_p M_l]  = \delta_{pl} \}$. We can use these
matrices to write any particular solution of the linear system $\mathrm{Tr}[\rho M_l ] = z_l$ as 
$\rho^{part} = \sum_{l=1}^L z_l \, \tau_l$. Combining this with (\ref{gensol}) we have 
\be
\label{moregensol}
\rho = \sum_{l=1}^L z_l \, \tau_l + \sum_{a=1}^{D_K} y_a  \mu^{(a)},
\ee
as the most general solution of (\ref{ineqconstr})
 provided that $z_l \in [m^{min}_l,m^{max}_l]$. If we apply the PPTSE 
criterion to (\ref{moregensol}) as before, we will obtain a linear combination of matrices representing the PPT
symmetric extension of a state satisfying this set of equations. We just need to once again construct the required LMIs
to impose the positive semidefiniteness of the extension and its partial transposes, and solve the resulting
SDP satisfying the constraints $z_l \in [m^{min}_l,m^{max}_l]$. These constraints can be 
imposed by another LMI, namely $\mathbf{diag}(z_1 - m^{min}_1,m^{max}_1 - z_1,\ldots,z_L - m^{min}_L,m^{max}_L - z_L) \succeq 0$,
showing that constraining the range of the variables does not change the SDP structure.

\subsection{Alternative SDP formulation}

We can formulate the search for a PPTSE as a slightly different SDP that has the advantage of performing
better numerically and providing a connection with entanglement measures (as discussed in Section \ref{entmeasures}). We will
replace the feasibility SDP discussed above by the following:
\begin{eqnarray}
\label{sdpalt}
{\mathrm{minimize}} &\ \  t \nonumber \\
{\mathrm{ subject \ to}} &\ \  F_0 + \sum_i x_i F_i + t \mathbf{1} \succeq 0.
\end{eqnarray}
Here, we have added a term proportional to the identity $\mathbf{1}$
to the affine combinations of the $F_i$ matrices,
and we minimize its coefficient $t$. The purpose of this is to make the SDP feasible: the LMI
can always be satisfied if we choose $t$ large enough. This makes the SDP solvers 
perform better in practice. To connect this SDP with the feasibility problem, we just need to realize that
$ F_0 + \sum_i x_i F_i \succeq 0$ is feasible if and only if $t_{opt} \leq 0$. If the optimal
value of $t$ is positive and bounded away from zero, the original SDP is infeasible
and the state does not have a PPTSE (and hence it is entangled). 

\subsection{Extension to the multipartite case}

Even though the approach described in this section considers only the bipartite case, the technique
can be extended to the multipartite case. In~\cite{doherty2005a} it was shown that requiring the existence
of PPTSE to any number of copies of any subset of parties of a multipartite state, gives a complete characterization
of the set of fully separable states. Again, the search for these extensions can be cast as an SDP, and failure
to find one implies entanglement of the state. The PPTSE algorithm was used to show
entanglement of a $2\otimes 2 \otimes 2$ state that has the property of being separable under any bipartition.
The algorithm presented here can thus also detect multipartite entanglement, although it will not
distinguish between inequivalent forms of multipartite entanglement (like $W$ and $GHZ$ entanglement in the case
of three qubits).

\section{Entanglement witnesses }
\label{witness}

Another useful feature of the PPTSE criterion is that, if the primal SDP is infeasible (i.e., there is
no separable state satisfying the constraints), the dual SDP provides a certificate of this fact in the 
form of an entanglement witness~\cite{doherty2003d}. Let us recall that an entanglement witness (EW) for a state
$\rho$ is a Hermitian operator $W$ that satisfies
\be
\mathrm{Tr}[\sigma_{sep} W] \geq 0 \ \ \mathrm{and} \ \ \mathrm{Tr}[\rho W] < 0,
\ee
where $\sigma_{sep}$ is {\it any} separable state.
These operators provide a proof of the entanglement of a given state. Entanglement witnesses are a consequence 
of the separating hyperplane theorem (or Hahn-Banach theorem) of convex geometry: if two
closed convex sets are disjoint and one of them is compact, there is a hyperplane that separates them.
In the context of checking separability of linearly constrained states,
the convex sets in question are the set of separable states and the affine subspace spanned by
all the solutions of the linear system (\ref{linear constraints}) (see Figure 1). If these two sets are disjoint, it means
that no separable state satisfies (\ref{linear constraints}); on the other hand, the separating hyperplane
theorem assures us that there is an entanglement witness that can certify the entanglement of every state
of the form (\ref{gensol}).

The dual SDP to (\ref{sdpalt}) takes the form
\begin{eqnarray}
\label{dualalt}
{\mathrm{maximize}} &\ \  -\mathrm{Tr}[F_0 Z]   \nonumber \\
{\mathrm{ subject \ to}} &\ \  Z \succeq 0 \nonumber \\
 &\ \   \mathrm{Tr}[F_i Z] = 0  \nonumber \\
 &\ \   \mathrm{Tr}[Z] = 1.
\end{eqnarray}
In~\cite{doherty2003d} it was shown how to use the solution of the dual SDP to construct an EW.
Without going into a detailed derivation we can point out the main elements of the correspondence
between the operator $Z$ and the corresponding EW. First, note that $F_0$ lies in a vector space
that is the direct sum of three copies of the space ${\cal H}_A^{\otimes 2} \otimes {\cal H}_B $ , so the operator
$Z$ lies in the same space. From (\ref{rhoextfix}) we can see that $F_0$ is a linear function of the 
matrix $\rho^{part}$, so we can write $F_0 = \Lambda(\rho^{part})$ for some linear map $\Lambda$
that operates on matrices in ${\cal H}_A \otimes {\cal H}_B $. If $\Lambda^\ast$ is the adjoint map,
we can write $\mathrm{Tr}[F_0 Z] = \mathrm{Tr}[\Lambda(\rho^{part})Z] = \mathrm{Tr}[\rho^{part} \Lambda^\ast (Z)] $,
so the objective of the dual SDP is minimizing the expectation value of the operator $\Lambda^\ast (Z)$ on the matrix
$\rho^{part}$. By applying the same line of reasoning, it is not difficult to see that the constraints 
$ \mathrm{Tr}[F_i Z] = 0$ in (\ref{dualalt}) allows us also to write
$\mathrm{Tr}[F_0 Z] =  \mathrm{Tr}[(\rho^{part} + \sum_{a=1}^{D_K} y_a  \mu^{(a)})\Lambda^\ast (Z)] $. This is
the expectation value of $\Lambda^\ast (Z)$ \emph{on all states compatible with} (\ref{linear constraints}). The
other constraints in (\ref{dualalt}) can be used to show that $\tilde{Z} = \Lambda^\ast (Z)$ is actually
positive on all pure product states as required for an EW~\cite{doherty2003d}. The dual SDP can be interpreted as minimizing
the expectation value of $\tilde{Z}$ on $\rho^{part} + \sum_{a=1}^{D_K} y_a  \mu^{(a)}$ over a particular
subset of EWs.
\begin{figure}
\includegraphics[scale=0.7]{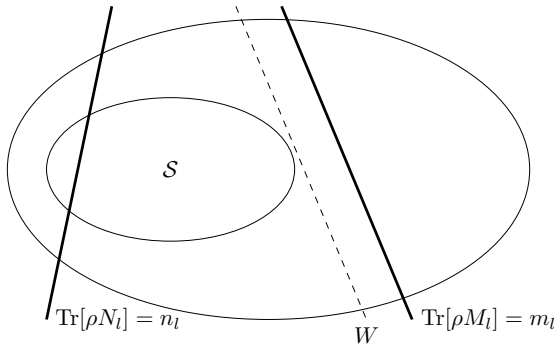}
\caption{\label{Fig:ew} The affine subspace defined by the linear constraints, which can either intersect the set of separable states ${\cal S}$ or not.
In the former case, there are separable states compatible with the constraints so no conclusion can be drawn about the
entanglement of the state. In the latter case, all states compatible with the constraints are entangled, and an entanglement
witness $W$ exists that certifies this fact.}
\end{figure}
Figure 1 gives a simple pictorial representation of the basis for this technique.
The key point in our case, where the
state is only partially determined, is that if the affine space defined by ($\ref{gensol}$) does not
intersect the set of separable states (i.e., all such states are entangled), the Hahn-Banach theorem guarantees
the existence of an EW that separates 
the set of separable states from \emph{every state in this affine subspace}. The dual SDP is used to construct one
such EW. Consequently, this approach is
not plagued by the ``fake entanglement" problem~\cite{horodecki1999c} that can arise when using the maximum entropy
method to infer the most probable state associated with (\ref{linear constraints}).

\section{Lower bounds on entanglement measures}
\label{entmeasures}

In the case where we are able to prove entanglement using the PPTSE criterion, we can use the
output of both the primal and dual SDPs to provide lower bounds on certain entanglement measures
and other related quantities. Consider the primal problem (\ref{sdpalt}) and let $t_{opt}$ be the optimal value.
If $t_{opt} > 0$ then all the states are entangled. But then $d_A^2 d_B t_{opt}$ is a lower bound on the 
minimum amount of the maximally
mixed state we need to add to a state satisfying (\ref{linear constraints}) to make it separable (in $2\otimes 2$ and 
$2\otimes 3$ this bound is tight). This is known as the random robustness of entanglement~\cite{vidal1999a}, $R_r(\rho)$, and quantifies
how robust the entanglement is against white noise. It also provides a lower bound on a geometric measure of 
quantum discord~\cite{debarba2012a}. 

The entanglement witness constructed from
the dual SDP can also be used to quantify the entanglement of the states satisfying (\ref{linear constraints}). Any entanglement
measure that can be expressed as 
\be
E(\rho) = \mathrm{max} \{0, -\min_{W\in\cal{M} }  \mathrm{Tr}[W \rho] \}
\ee
with ${\cal M}$ a subset of entanglement witnesses, is
referred to as \emph{witnessed entanglement}~\cite{brandao2005a}. The set $\cal{M}$ determines which 
particular measure this expression represents. Several 
well-know measures are of this form, such as the best separable approximation $BSA(\rho)$, the negativity ${\cal N}(\rho)$, 
and the concurrence ${\cal C}(\rho)$.
Clearly, any $W \in \cal{M}$ that satisfies $\mathrm{Tr}[W \rho] <0$ provides a lower bound to $E(\rho)$. In particular,
the quantities 
\be
E_{n,m}(\rho) = \mathrm{max} \{0,  -\min_{W\in{\cal{M}}_{n,m} } \mathrm{Tr}[W \rho] \}
\ee
($n,m \geq 0$) with ${\cal{M}}_{n,m} = \{ W  : -n\mathbf{1} \preceq W \preceq m \mathbf{1}\}$ a subset of entanglement witnesses,
 are entanglement
monotones, and satisfy $E_{n,m}(\rho) \to n BSA(\rho)$ when $m\to \infty$, where $BSA(\rho)$ is the best separable
approximation to $\rho$~\cite{brandao2005a}. Since $E_{n,m}(\rho)$ is obviously monotonically increasing with $m$ (for fixed $n$) 
and any entanglement witness  $W$ 
must be in some ${\cal{M}}_{n,m}$,
$\mathrm{Tr}[W \rho]$ provides a lower bound on $BSA(\rho)$, which is an entanglement measure. 
This analysis is
just an illustration of the connection between the PPTSE criterion and entanglement measures, and does not pretend
to give the best bounds possible.

\section{Example}
\label{example}

Let us use a simple example to illustrate the power of this approach. Consider a system that produces two photons
and we want to determine if they are entangled in the polarization basis. One possible approach is
to do quantum state tomography. This can be accomplished by measuring the 16 observables~\cite{james2001a}  given by
$\hat{\mu}_i \otimes \hat{\mu}_j \, (i,j=0,1,2,3)$ with 
\begin{eqnarray}
\hat{\mu}_0 &=& |H\rangle\langle H| +|V\rangle\langle V| \nonumber \\
\hat{\mu}_1 &=& |H\rangle\langle H| \nonumber \\
\hat{\mu}_2 &=& |D\rangle\langle D| \nonumber \\
\hat{\mu}_3 &=& |R\rangle\langle R|
\end{eqnarray}
with $|D\rangle =(|H \rangle - |V \rangle) / \sqrt{2}$ and 
$|R\rangle =(|H \rangle - i |V \rangle) / \sqrt{2}$. Note that these operators are all positive on pure product states
and so they are good candidates to be entanglement witnesses. Assume that we measure these observable and we obtain
\begin{eqnarray}
\label{evs}
0.48 \leq & \mathrm{Tr}[(\hat{\mu}_1 \otimes \hat{\mu}_1) \rho] & \leq 0.5 \nonumber \\
0.24 \leq & \mathrm{Tr}[(\hat{\mu}_1 \otimes \hat{\mu}_2) \rho] & \leq 0.25 \nonumber \\
0.48 \leq & \mathrm{Tr}[(\hat{\mu}_2 \otimes \hat{\mu}_2) \rho] & \leq 0.5 \nonumber \\
0 \leq & \mathrm{Tr}[(\hat{\mu}_3 \otimes \hat{\mu}_3) \rho] & \leq 0.02.
\end{eqnarray}
Note that all expectation values are non negative, so they cannot show entanglement by themselves.
However, applying our test we find that there is an entanglement witness given by
\bea
Z & = & 0.1343 |HH\rangle \langle HH| + 0.3977 |HV\rangle \langle HV| + \nonumber \\
& & 0.234 (|VH\rangle \langle VH| +|VV\rangle \langle VV| ) + \nonumber \\
& & + \{(0.0658 + i 0.1583) (|HH\rangle \langle VH| 
+|HV\rangle \langle VV| + \nonumber \\
& & +|VH\rangle \langle VV|) + h.c. \} + \nonumber \\
& & +  \{ -0.2242 |HH\rangle\langle VV| +
0.0925 |HV\rangle \langle VH| + h.c \} \nonumber \\
\eea
such that $\mathrm{Tr}[Z \rho] < - 0.0168$ for all states
satisfying the constraints (\ref{evs}). Moreover, $Z \in {\cal{M}}_{1,1}$,  so this result provides a lower bound
on the best separable approximation, i.e., $BSA(\rho) \geq 1.68 \times 10^{-2}$. The primal SDP also 
computes a lower bound on the random robustness, $R_r (\rho) \geq 8 \times 0.0168 = 0.1344$.  Additional information
about the state can improve these bounds. For example, if we add $\mathrm{Tr}[(\hat{\mu}_1 \otimes \hat{\mu}_3) \rho] 
\in [0.24,0.25]$ to the constraints we now obtain a new entanglement witness $Z'$ such that
$\mathrm{Tr}[Z' \rho] < - 0.021$, which translates to $BSA(\rho) \geq 2.1 \times 10^{-2}$, and 
$R_r (\rho) \geq 0.168$ (the MATLAB code used is available online from the author~\cite{CETcode}).

\section{Some important features of the approach}
\label{features}

Having described the idea behind this method for detecting entanglement from partial state information,
we can now shift our attention to more general features regarding its usefulness and limitations. First,
we want to stress an important feature of this technique: independently of the number of constraints available,
if we are allowed enough computational resources we are guaranteed to arrive at a definite answer to the 
question ``Are all states satisfying these constraints entangled, or is there  at least one such state that is separable?".
This is accomplished by applying successive steps in the PPTSE hierarchy and its dual. Furthermore, if
the answer is that all such states are entangled, this affirmation is free of the ``fake entanglement" issue
that appears in maximum entropy inference approaches.

The second question that arises is whether there is a clear correlation between the number of constraints
(i.e., the amount of information about the state) and the number of steps in the PPTSE hierarchy and its dual
we need to apply. One could naively expect that if the number of constraints is very small compared to the 
number of parameters in the density matrix, there would have to be some separable state that satisfies them.
However, it is easy to see that this is not the case: if the partial information is the 
expectation value of a single observable that happens to be an entanglement witness, and its value is negative, we
know for sure that the state is entangled. 

Let us assume that a set of constraints is shown to be compatible with a separable state. One may consider the
question of which extra observables we should add to further constraint the state and determine its entanglement.
Without extra assumptions, this question does not have a definite answer: if the unknown state
is actually separable, no matter which observables we choose, their expectation values will always be
compatible with a separable state. On the other hand, if a set of constraints is sufficient to prove entanglement
and provides lower bounds on some entanglement measures, an extra constraint may be useful to 
improve that bound and so it may be worth performing that extra measurement. 
If for some reason we had some additional information, for example, 
the state is being drawn from some distribution, we may use it to find an observable that maximizes the probability
of detecting the entanglement (if it exists). But the approach considered in this paper is aimed
precisely at the situation where we do not have any more information than the one provided
by the linear constraints associated with the measured observables.

From this discussion we can start to form a picture of when this  approach would fail in practice, i.e., it would not produce a definite answer
after a reasonable number of steps. If we look at the PPTSE part of the method (proving entanglement), we can see that
if the affine space defined by (\ref{gensol}) intersects sets ${{\cal S}}_p^N $ with $N$ large (${{\cal S}}_p^N $ is the
set of states with PPT symmetric extensions to $N$ copies of $A$), even if the states are actually entangled
it will take at least $N$ steps to prove this fact. On the other hand, if we consider the dual approach and (\ref{gensol}) 
intersects the set of separable states but does not intersect ${\tilde{\cal S}}_p^N $ for $N$ small, the procedure will take a
long time to provide the required separable decomposition. Since ${{\cal S}}_p^N $ and ${\tilde{\cal S}}_p^N $ approach
${\cal S}$ from the outside and the inside respectively, the most challenging situation for this procedure
occurs when (\ref{gensol}) is ``almost" tangent  to the set of separable states, either intersecting it or not. 
For a given instance, if this approach fails
 to provide an answer in a reasonable number of steps, adding extra constraints may be of some help (although this is
 not guaranteed in general). 

\section{Conclusions}
\label{conclusions}

In this article we have introduced a sequence of tests that can determine the entanglement or
separability of a state when only partial information is available. Using a set of linear constraints
on the density matrix, such as the ones associated with the expectation values of a set of observables,
we can apply the PPTSE separability criterion and its dual to determine whether all the states satisfying
these constraints are entangled or if there is one such state that is separable. When entanglement is proven
by this method, the algorithm constructs an entanglement witness that can be shown to certify the
entanglement of all states that satisfy the constraints. On the other hand, when a separable state is
found that is compatible with the constraints, a separable decomposition is also constructed to 
prove this fact.

Even though this approach is technically very similar to the original PPTSE criterion, its range
of applicability is radically different. The original PPTSE criterion requires as input the complete
density matrix of the state that we want to analyze. This is very useful for theoretical considerations,
when the state is explicitly constructed to accomplish some particular task (such as some communication
protocol or a particular scheme for quantum key distribution), but it is not as helpful
when the state comes from actual experimental measurements on a physical system. Before applying
the PPTSE criterion in this case the state must be reconstructed using a procedure like quantum state tomography,
but this has the disadvantage of typically requiring a large number of measurements and it is not even guaranteed
to provide a consistent answer. In contrast, the sequence of tests in this paper can be applied directly to experimental
data, and in the case where entanglement if proven it also provides lower bounds on entanglement measures
and other related quantities.

The method introduced here avoids performing quantum state tomography and analyzes what can be said about
the entanglement of the state using only the partial information provided. When this information is in the form of a
set of linear constraints on the elements of the density matrix, the state in question belongs to an affine subspace in the
space of density matrices. If this affine subspace does not intersect the set of separable states (see Figure 1), the Hahn-Banach
theorem guarantees the existence of a separating hyperplane that is associated with an entanglement witness that certifies
entanglement for all states in such an affine subspace. The dual SDP of the criterion presented here can be interpreted
as a search for such a separating hyperplane over a restricted set of entanglement witnesses. The completeness of
the PPTSE criterion can be extended to this case to guarantee that if all states compatible with the constraints
are entangled, such an entanglement witness will be found. On the other hand, if the subspace intersects
the set of separable states the dual test will eventually provide a state in that intersection. These two 
central features imply that it is not possible for the method to certify entanglement if there is a single
separable state that satisfies the constraints. Thus, this approach is free from the ``fake entanglement" issue 
common to maximum entropy inference based methods. Given that experimental data can be used as input
to these tests with basically no preprocessing, this technique could be a very useful and practical tool
for experimentally certifying entanglement of real physical systems.

\section{Acknowledgments}
I would like to thank Sergio Boixo, Daniel Lidar and Pablo Parrilo for many useful conversations and suggestions
that helped to refine the approach presented in this work.

\bibliographystyle{prsty}
\bibliography{FS-Bibliography}

\begin{thebibliography}{10}

\bibitem{nielsen2000}
M.~N. Nielsen and I.~L. Chuang, {\em Quantum computation and quantum
  information} (Cambridge University Press, Cambridge, 2000).

\bibitem{gurvits2003a}
L. Gurvits,  in {\em STOC '03: Proceedings of the thirty-fifth annual ACM
  symposium on Theory of computing} (ACM Press, New York, NY, USA, 2003), pp.\
  10--19.

\bibitem{vogel1989a}
K. Vogel and H. Risken, Phys. Rev. A {\bf 40},  2847  (1989).

\bibitem{doherty2002a}
A.~C. Doherty, P.~A. Parrilo, and F.~M. Spedalieri, Phys. Rev. Lett. {\bf 88},
  187904  (2002).

\bibitem{navascues2009a}
M. Navascu\'es, M. Owari, and M.~B. Plenio, Phys. Rev. Lett. {\bf 103},  160404
   (2009).

\bibitem{doherty2003d}
A.~C. Doherty, P.~A. Parrilo, and F.~M. Spedalieri, Phys. Rev. A {\bf 69},
  022308  (2004).

\bibitem{navascues2009b}
M. Navascu\'es, M. Owari, and M.~B. Plenio, Phys. Rev. A {\bf 80},  052306
  (2009).

\bibitem{doherty2005a}
A.~C. Doherty, P.~A. Parrilo, and F.~M. Spedalieri, Phys. Rev. A {\bf 71},
  032333  (2005).

\bibitem{horodecki1999c}
R. Horodecki, M. Horodecki, and P. Horodecki, Phys. Rev. A {\bf 59},  1799
  (1999).

\bibitem{vidal1999a}
G. Vidal and R. Tarrach, Phys. Rev. A {\bf 59},  141  (1999).

\bibitem{debarba2012a}
T. Debarba, T.~O. Maciel, and R.~O. Vianna, Phys. Rev. A {\bf 86},  024302
  (2012).

\bibitem{brandao2005a}
F.~G. S.~L. Brand\~ao, Phys. Rev. A {\bf 72},  022310  (2005).

\bibitem{james2001a}
D.~F.~V. James, P.~G. Kwiat, W.~J. Munro, and A.~G. White, Phys. Rev. A {\bf
  64},  052312  (2001).

\bibitem{CETcode}
\mbox{http://www.isi.edu/people//fspedali/entanglement{\_}code}.

\end{thebibliography}

\end{document}